 \documentclass[final,5p,times,twocolumn]{elsarticle}

\usepackage{amssymb}
\usepackage{lipsum}
\usepackage[hidelinks]{hyperref}
\setcitestyle{square}
\journal{NIM A}

\begin{document}

\begin{frontmatter}



\title{Discharge quenching mechanism and performance of RPWELL with tunable 3D printed resistive plates, charge evacuation in semiconductive glass RPWELL and discharge quenching for Cryogenic-RWELL over a wide range of resistivity}


\author[a]{Abhik Jash}
\author[a]{Luca Moleri}\ead{luca.moleri@weizmann.ac.il}
\author[a]{Andrea Tesi}
\author[a]{Shikma Bressler}

\affiliation[a]{organization={Weizmann Institute of Science},%Department and Organization
            %addressline={}, 
            city={Rehovot},
            %postcode={}, 
            %state={},
            country={Israel}}
            
\begin{abstract}
Resistive electrodes are used in gaseous detectors to quench electrical discharges. This helps to protect delicate electrodes and readout electronics and to improve the stability of the detector operation. An RPWELL is a THGEM-based WELL detector with a resistive plate coupled to a conductive anode. Till now, the choice of the resistive plate was limited to a few materials, like LRS Glass and Semitron. These materials have fixed resistivities and, sometimes, thickness and area limitations. This restricts the potential usage of the detector to a rather small range of applications, as well as the possibility of studying in depth the physics processes governing the discharge quenching mechanism.
In our present study, we used a new plastic material doped with carbon nanotubes to produce resistive plates with a commercial 3D printer. This method has the flexibility to produce samples of different thicknesses and different resistivity values. We describe here the sample production and characterize the RPWELL performance with different resistive plates. In particular we show the dependence of discharge quenching on the thickness and resistivity of the plate. The dynamics of the charge carriers in the material is proposed as an explanation for the long gain recovery time after a discharge.
\end{abstract}




\end{frontmatter}

%\tableofcontents



\section{Introduction}
\label{introduction}
Thick gaseous electron multipliers (THGEM) are employed in various detector concepts and applications (for a recent review see \cite{bressler2023thick}). In THGEM detectors, the electron avalanche multiplication in the gas occurs within a hole of $\sim$mm diameter, typically produced by standard printed circuit board techniques. As in most micro-pattern gaseous detectors (MPGD), the occurrence of energetic electric discharges through the gas is a severe limiting factor for THGEM operation. This problem can be mitigated by coupling a single-sided THGEM electrode to a highly resistive plate, in a resistive-plate WELL (RPWELL) configuration \cite{rubin2013first}. In the resistive material, the electron mobility is reduced with respect to that in a bare metallic anode. In case of a discharge, this translates into a local decrease of the electric field intensity and, consequently, quenching of the discharge energy. On the other hand, it also limits the detector's operation in high radiation environments. Ideally, an optimized resistivity value should be selected for each application in accordance with its specific operation conditions. However, the availability of electrostatic-dissipative (ESD) materials with a resistivity in the relevant range, 10$^{8}$-10$^{11}$$\Omega$cm, is very limited, e.g. Semitron ESD225\footnote{\href{https://www.mcam.com/en/products/shapes/advanced/semitron}{MGC Mitsubishi Chemical Group - Quadrant}} and semiconductive glass \cite{wang2008study}. Moreover, obtaining resistive plates with custom thickness is quite expensive and sometimes impossible. 

Using fused deposition modeling 3D printing provides new opportunities. We used a commercial filament of ABS loaded with multi-wall carbon nanotubes (CNT)\footnote{\href{https://www.3dxtech.com/product/3dxstat-esd-abs/}{3DXSTAT ESD ABS}} to tune the resistivity of the printed plate by changing the temperature settings of the printer's nozzle and bed plate during deposition~\cite{jane2022characterization}. We produced plates of different thickness and resistivity and operated them in an RPWELL configuration to investigate discharge quenching effects. Preliminary measurements of charge evacuation time in a semiconductive glass RPWELL and of discharge quenching in a Cryogenic-RWELL detector \cite{Tesi:2023ale} with variable resistivity are also reported.

\section{3D printed resistive plates}
\begin{figure}
	\centering 
	\includegraphics[width=0.25\textwidth, angle=-90]{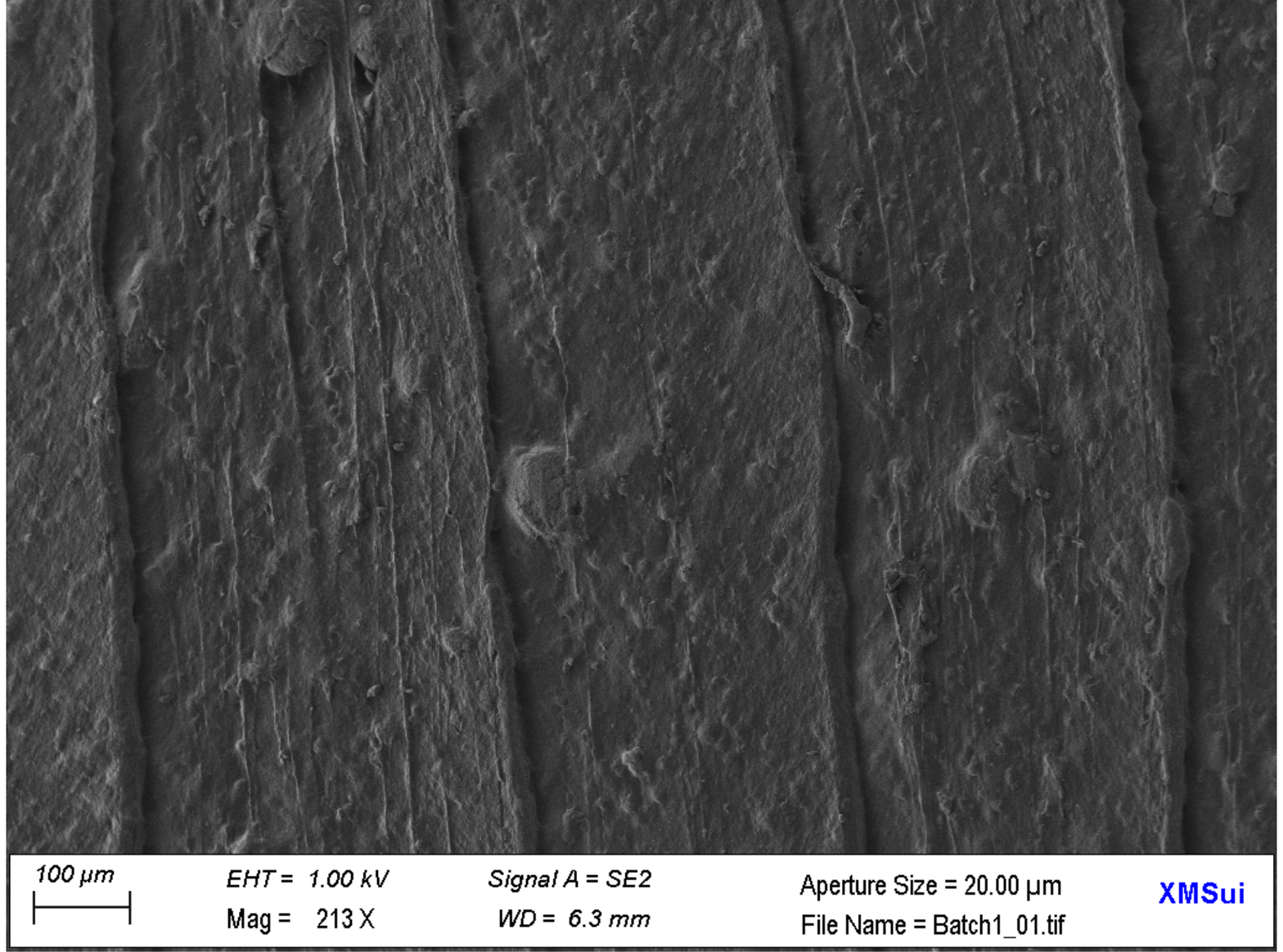}	
 \includegraphics[width=0.25\textwidth, angle=-90]{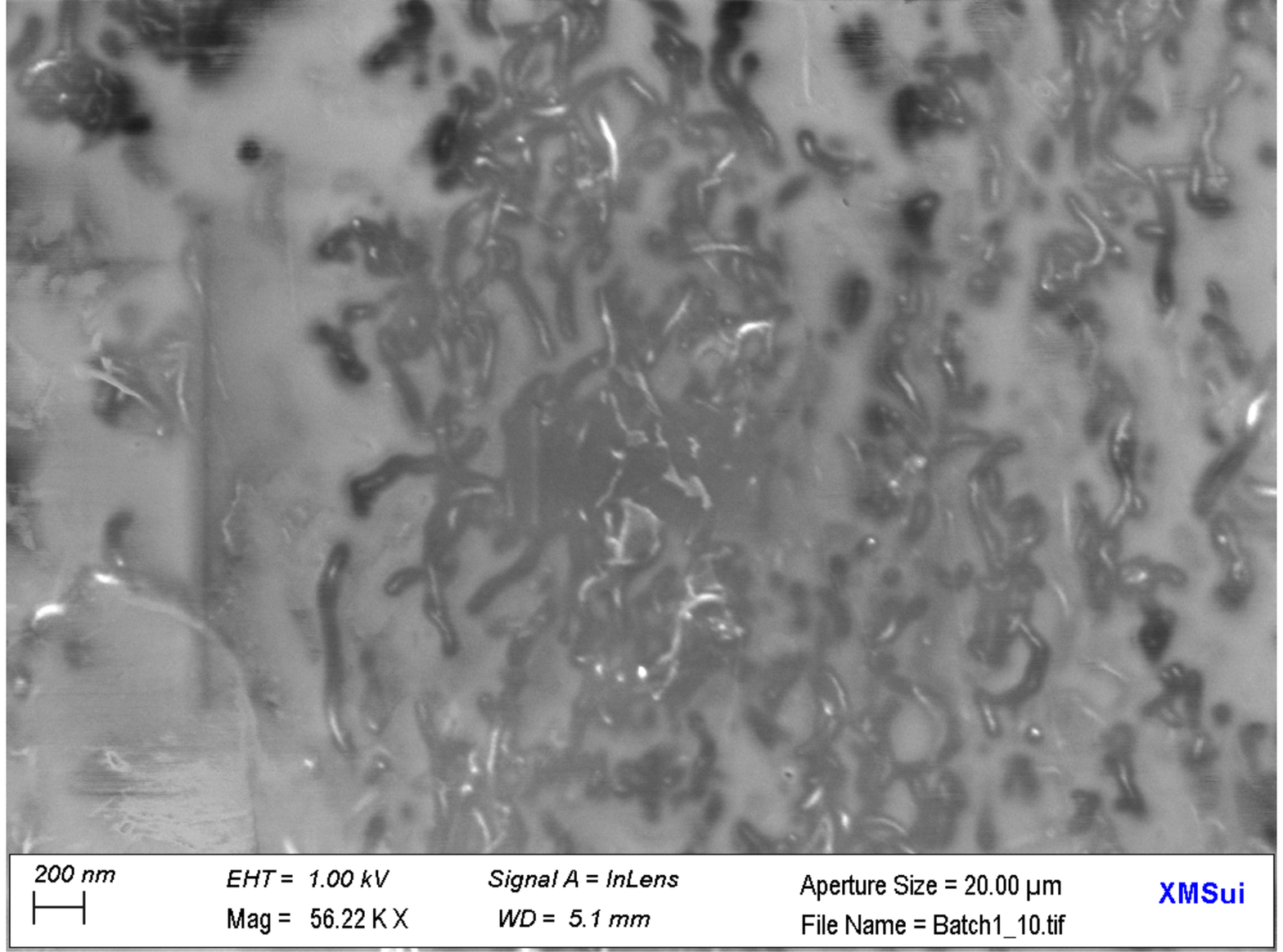}	
	\caption{SEM images of a 3D printed resistive plate. Top: detail of the filament deposition lines. Bottom: dispersion of CNTs in the ABS matrix.} 
	\label{fig: resistive plates}
\end{figure}

%%\label{}
A set of resistive plates was produced with different nozzle temperatures. The side attached to hot bed during the production had a smooth surface. As seen in the Scanning Electron Microscopy (SEM) image in figure~\ref{fig: resistive plates} (left), The other surface was rough due to the filament deposition pattern. In figure~\ref{fig: resistive plates} (right), the directionality of the CNT distribution in the plastic matrix are clearly visible.

\begin{figure}
	\centering 
	\includegraphics[width=0.40\textwidth]{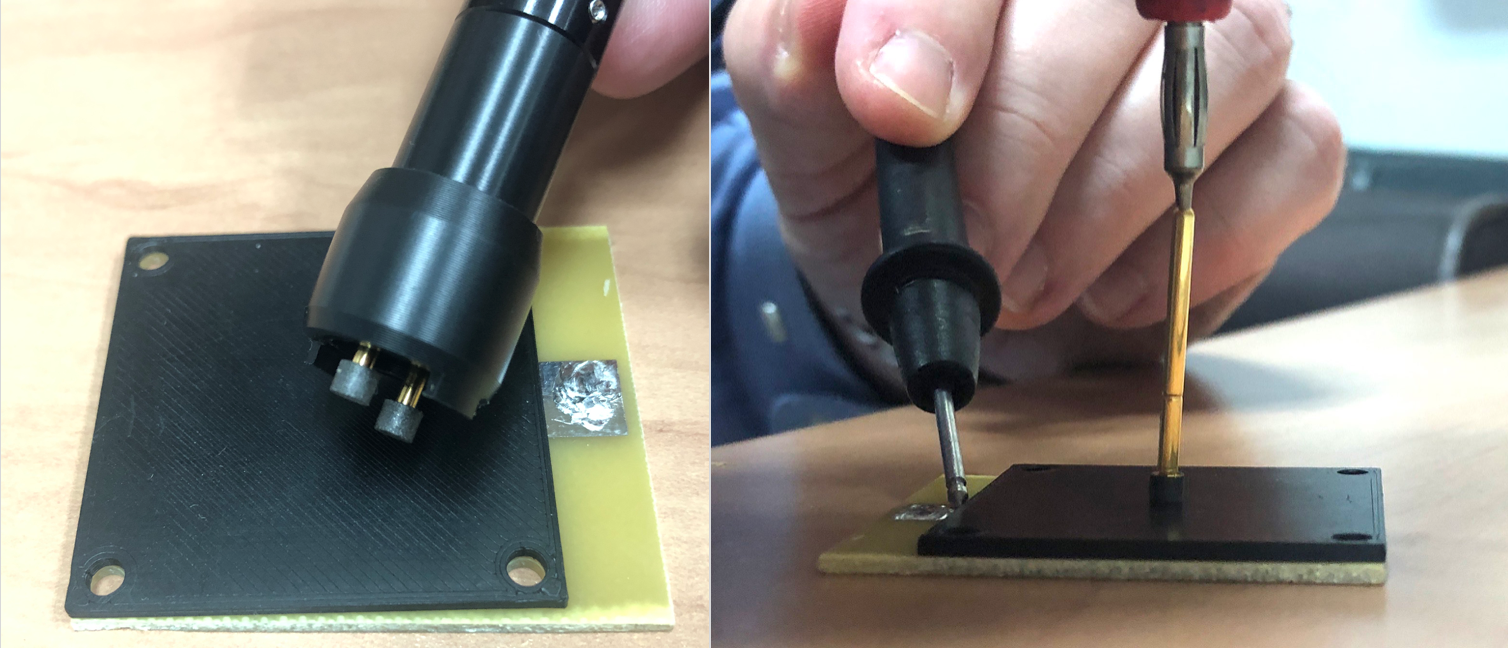}	
	\caption{Measurement of surface (left) and bulk (right) resistance of resistive plate anodes.} 
	\label{fig: plates measurement}%
\end{figure}

\begin{figure}
	\centering 
	\includegraphics[width=0.45\textwidth]{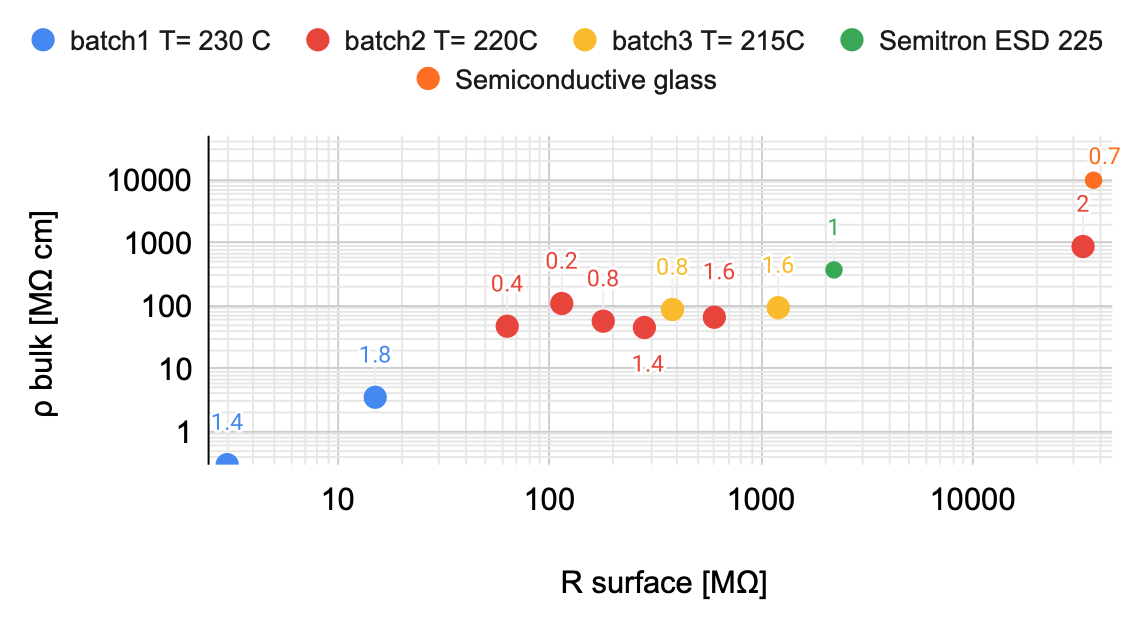}	
	\caption{Bulk resistivity and surface resistance of the produced ABS resistive plates. Semitron ESD225 and semiconductive glass were also measured for reference. The labels indicate the thickness of each plate in mm.} 
	\label{fig: plates characterization}%
\end{figure}

The rough side of the plates was attached to a plain copper anode with a conductive transfer tape\footnote{\href{https://www.3m.com/3M/en_US/p/d/b00043204/}{3M Electrically Conductive Adhesive Transfer Tape 9707}}. The smooth side faced the gas in the RPWELL assembly. Its surface resistance, $R_s$, was measured using a two point probe\footnote{\href{https://www.prostatcorp.com/product/prf-922b-miniature-two-point-probe}{Prostat PRF-922B}} and the bulk resistance, $R_b$, between a single pin of the probe and the anode was measured with an insulation tester\footnote{Fluke 1587} (see figure~\ref{fig: plates measurement}). The bulk resistivity value was calculated as $\rho = R \cdot D / t$, where R is the measured resistance, $D$ is the area of the probe and $t$ is the plate thickness in mm. The results in figure \ref{fig: plates characterization} show a large variability of surface resistance from sample to sample even if produced at the same temperature values. The thickness of each plate is marked by a label.

\section{RPWELL detector characterization}

Each of the resistive anodes was, in turn, assembled in the same RPWELL configuration. The THGEM electrode was 0.8~mm thick, with 0.5~mm diameter holes distributed in a hexagonal pattern with 1~mm pitch. A rim of 0.1~mm was etched around the holes. The RPWELL detector was characterized in terms of effective gain and electrical breakdown properties in the same setup and with the same methodology described in \cite{jash2022electrical}. The operation gas was Ar:CO$_2$ (93:7). For each configuration, the maximum operation voltage was determined by the appearance of discharges or quenched discharges, causing a significant distortion of the x-ray spectrum shape.

\subsection{Effective gain}

\begin{figure}
	\centering 
	\includegraphics[width=0.37\textwidth]{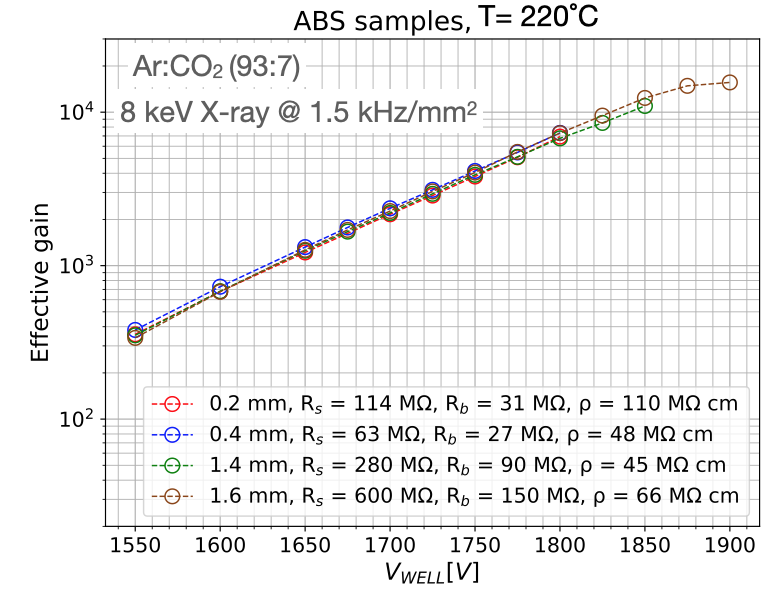}	
 \includegraphics[width=0.37\textwidth]{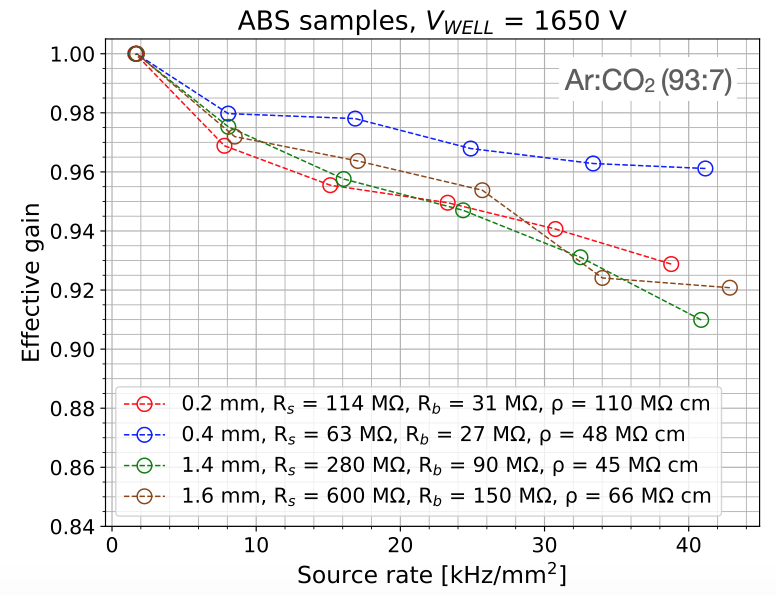}	
	\caption{Left: RPWELL effective gain curves for resistive plates with similar resistivity and different thickness. Right: For the same anode plates, effective gain in source rate scan at a fixed voltage.} 
	\label{fig: gain_vs_t}%
\end{figure}

In the first set of measurements, we compared the effective gain curve of configurations with similar surface and bulk resistance (in the ranges 114~M$\Omega <$  R$_s <$ 600~M$\Omega$ and 30~M$\Omega <$ R$_b <$ 150~M$\Omega$) and large thickness variation (0.2-1.6 mm). The results are shown in figure~\ref{fig: gain_vs_t}. As can be seen, the plate's thickness does not affect the value of the effective gain. This suggests that, at these resistivity values, the plates behave like a conductor; don't change the weighting field, and thus don't affect the signal induction \cite{jash2022electrical}. 

The maximum gain achieved was $\sim$10$^4$, corresponding to a total of $\sim$10$^6$ electrons in agreement with the Raether limit \cite{jash2022electrical}. In general, higher gains were achieved for larger thickness, R$_s$ and R$_b$ values. The samples with higher thickness and resistance reached a saturated regime (slow rise with voltage) when exceeding the gain of $\sim$10$^4$. 

The gain drop as a function of source rate does not show a clear dependence on thickness or resistance in the current configuration. This might be different when irradiating the entire detector area instead of using a collimated source.

\begin{figure}
	\centering 
	\includegraphics[width=0.37\textwidth]{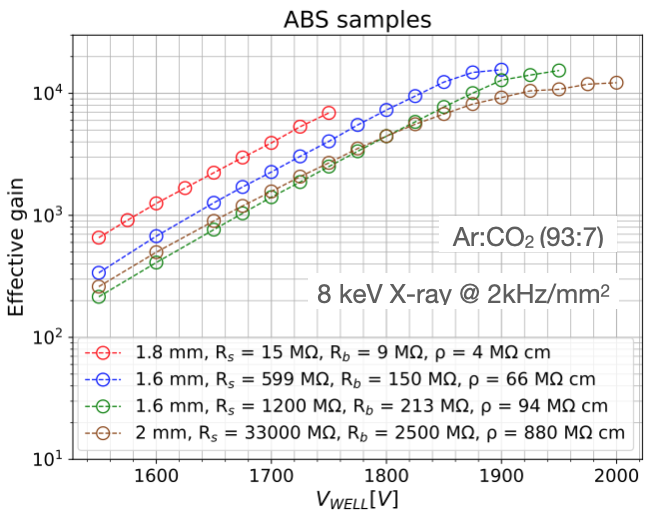}	
 \includegraphics[width=0.37\textwidth]{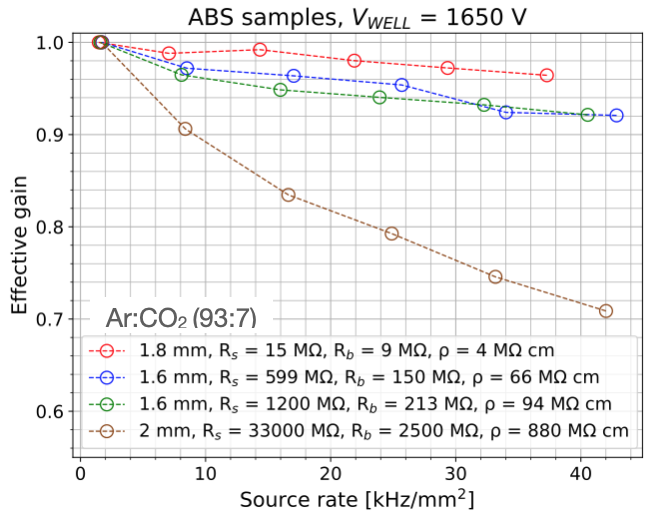}	
	\caption{Left: RPWELL effective gain curves for resistive plates with similar thickness and different resistance values. Right: For the same anode plates, effective gain in source rate scan at a fixed voltage.} 
	\label{fig: gain_vs_R}%
\end{figure}

In the second set of measurements we compared RPWELL configurations with resistive plates of a similar thickness and different resistance in the range 15~M$\Omega <$  R$_s <$ 33~G$\Omega$, 9~M$\Omega <$ R$_b <$ 2.5~G$\Omega$. The results are shown in figure~\ref{fig: gain_vs_R}. As can be seen, the effective gain curves shift to higher voltage values for increasing resistivity. It is also possible to observe a deviation of the gain trend from exponential, starting at around 1700~V and more pronounced for the most resistive sample. This effect hints to a local voltage drop due to the current flowing in the resistive plate, which might also cause the shift of the gain curves. The latter might also be due in part to a reduced weighting field as the material properties are more similar to those of a dielectric \cite{jash2022electrical}. A possible way to discriminate between the two phenomena would be recording the flowing currents to measure the absolute gain of the detector, which is not affected by the weighting field.
The gain drop as a function of source rate does not show a clear trend for the dependence on resistance, except for the most resistive sample. All the curves show a highly saturated gain regime towards the upper end of the dynamic range, except for the one related to the plate with lowest resistance. This points toward the existance of a transition region in which the resistivity becomes sufficient to allow operation in the presence of quenched discharges: 15 M$\Omega <$  R$_s <$ 600~M$\Omega$, 9~M$\Omega <$ R$_b <$  150~M$\Omega$, 4~M$\Omega \cdot$cm $< \rho <$ 66~M$\Omega \cdot$cm 

\subsection{Electric discharges}
Electric discharges in the RPWELL detector were monitored and characterized following the two methods described in \cite{jash2022electrical}: a) The current drawn from each electrode through the power supply was digitized and recorded. b) The induced signal on the anode was shaped by a current amplifier and the discharge intensity was obtained by integrating each current spike. This allows detecting quenched discharges that do not cause measurable currents. Characterizing RPWELL detectors with different resistive plates resulted in the identification of three different discharge regimes: no-quenching, transition and quenching regime. 

\begin{figure}
	\centering 	
 \includegraphics[width=0.5\textwidth]{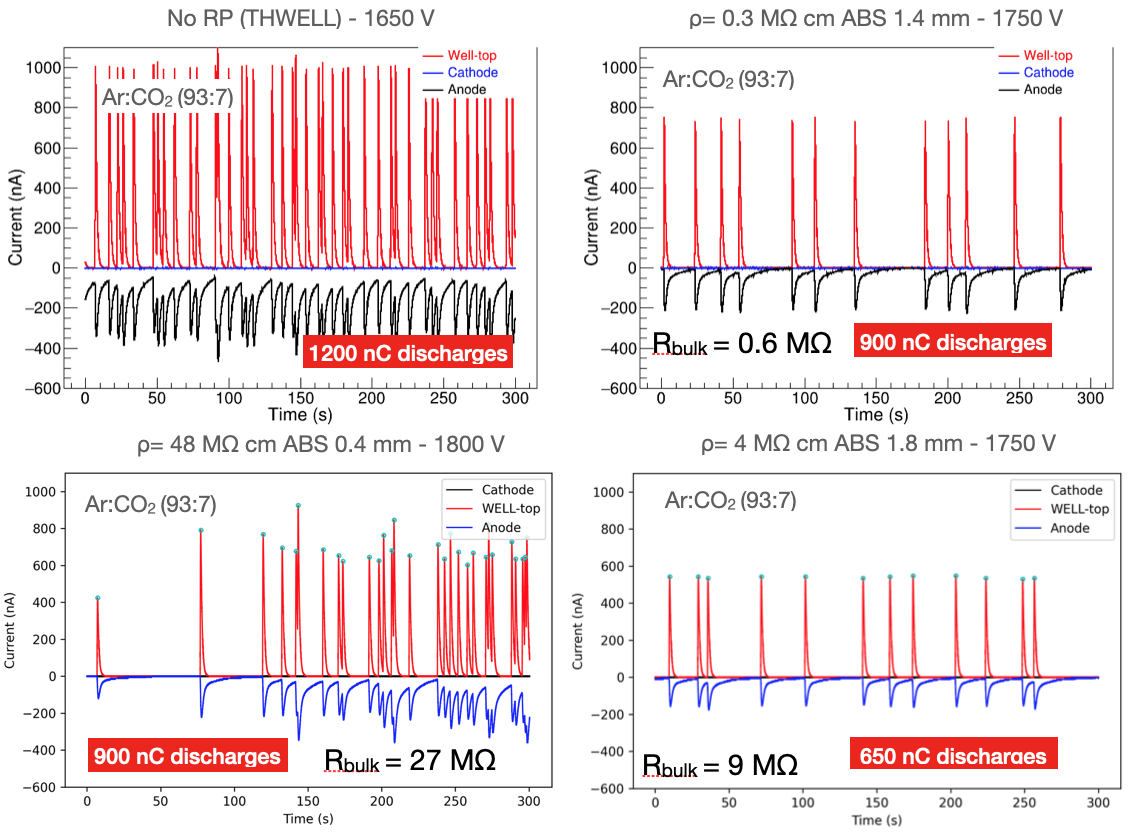}	
	\caption{Non-quenching discharge regime. Top-left: discharges in THWELL detector (no resistive plate). Top-right and bottom: discharges in RPWELL with different resistive plate thickness and resistivity.} 
	\label{fig: discharges_not_quenching}%
\end{figure}

\begin{figure}
	\centering 	
 \includegraphics[width=0.5\textwidth]{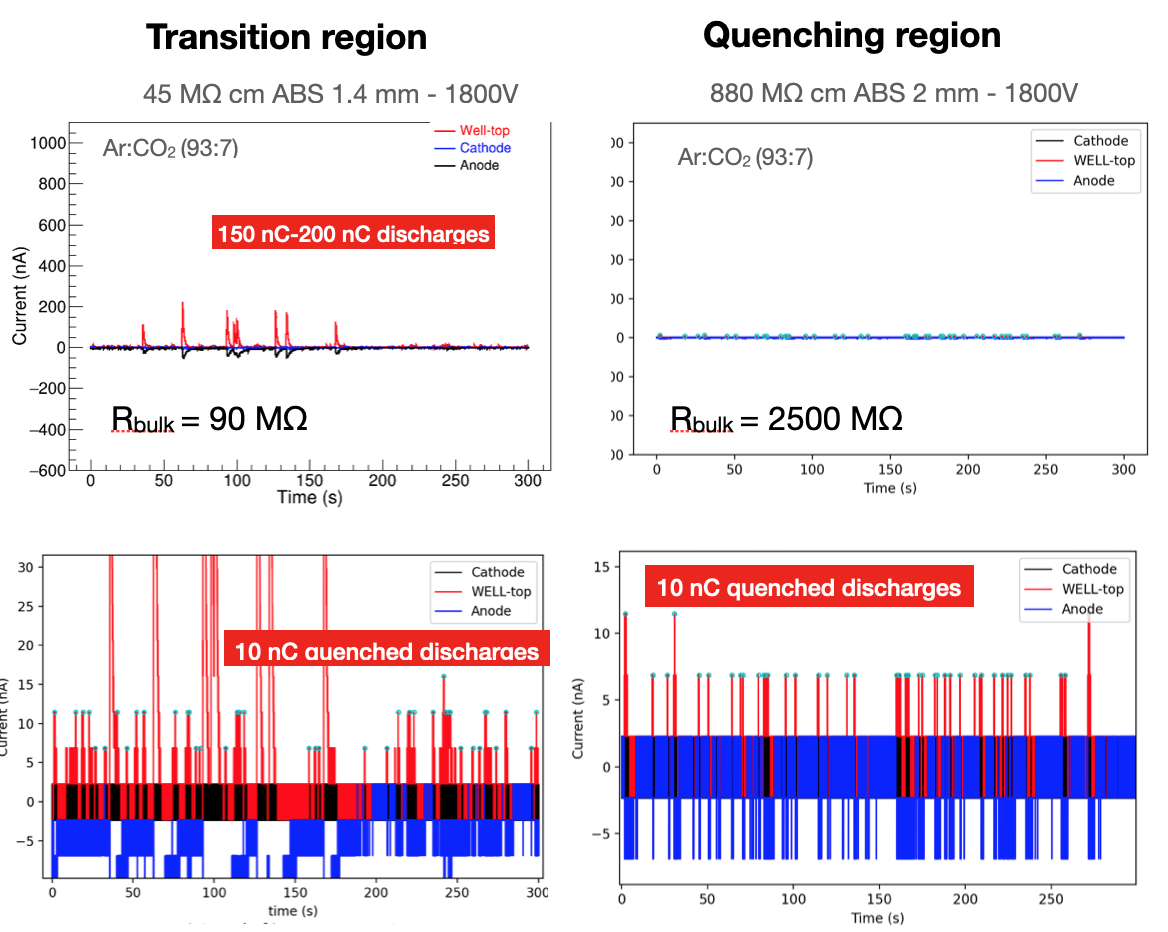}	
	\caption{Transition (left) and quenching (right) discharge regimes. Bottom: zoom in the y axis of the top plots.} 
	\label{fig: discharges_quenching}%
\end{figure}

A non-quenching behavior was observed for resistance values in the range R$_b <$ 50~M$\Omega$. Examples of discharges recorded from the power supply currents are shown in figure~\ref{fig: discharges_not_quenching}. Average intensity values are indicated in the figure.  A measurement from a THWELL detector (without resistive plate) is shown for reference. It can be seen that in this regime the intensity of the discharges is only mildly quenched by both thin (0.4~mm) and thick (1.8~mm) resistive plates, whereas the discharge probability is reduced with respect to that of a THWELL, for all the samples tested. Moreover, the difference between the 9~M$\Omega$ and the 27 M$\Omega$ samples suggests that the thickness might be important in the non-quenching discharge regime.

In the transition region, 50~M$\Omega <$ R$_b <$ 250~M$\Omega$ (examples in figure~\ref{fig: discharges_quenching}-left), some low intensity discharges ($\sim$200~nC) still appear from the power supply current monitor, together with quenched discharges ($\sim$10~nC) barely visible above noise. The extent of the region might be attributed to local non-uniformity and defects of the resistive plates. In the quenching region R$_b >$ 250~M$\Omega$ (examples in figure~\ref{fig: discharges_quenching}-right), only quenched discharges are present. 

\begin{figure}
	\centering 	
 \includegraphics[width=0.45\textwidth]{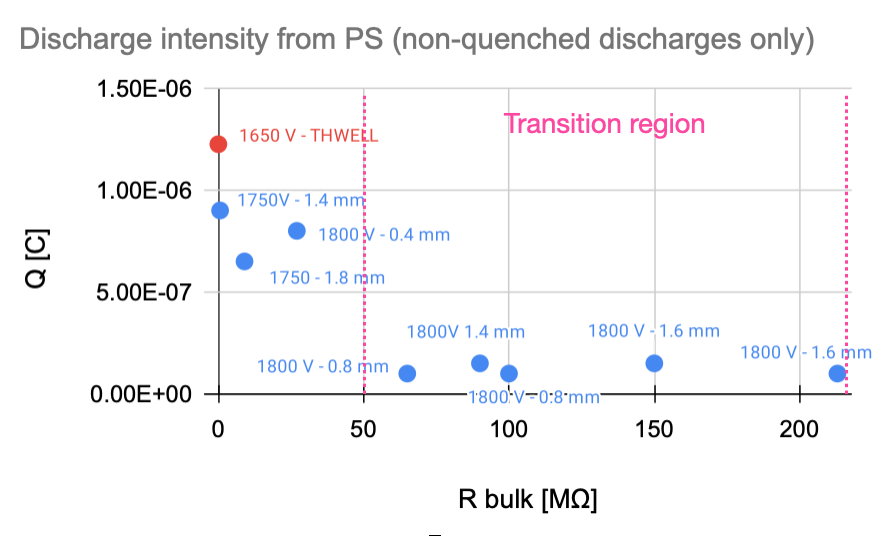}	
	\caption{Intensity of non-quenched discharges in RPWELL with all the tested resistive plates. The labels show the applied voltages and thickness.} 
	\label{fig: discharges_intensity_PS}%
\end{figure}

\begin{figure}
	\centering 	
 \includegraphics[width=0.5\textwidth]{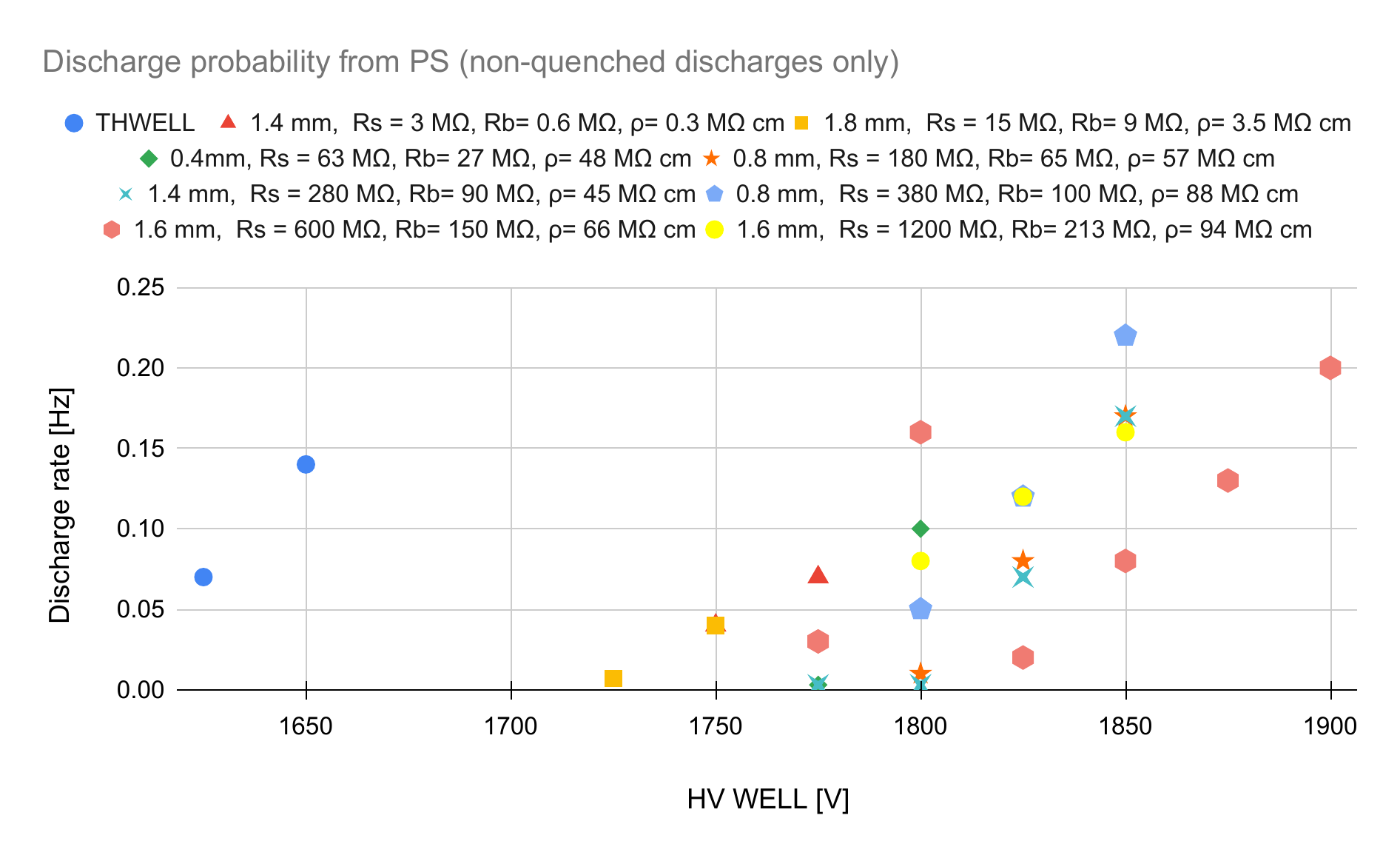}	
	\caption{Discharge rate of non-quenched discharges in RPWELL with all the tested resistive plates.} 
	\label{fig: discharges_probability_PS}%
\end{figure}

Figure~\ref{fig: discharges_intensity_PS} presents a summary plot of the intensity of non-quenched discharges in RPWELL with all the tested resistive plates. It can be seen that the discharges in the transition region are similar regardless the resistance value.
Figure~\ref{fig: discharges_probability_PS} shows the rate of non-quenched discharges in all the tested RPWELL configurations. The discharge rate is lower for any RPWELL configuration with respect to the bare THWELL (no resistive plate), even for the non-quenching resistive plates. Moreover, the discharge rate is similar for all resistive plates (considering that the gain is affected for the most resistive samples as shown in figure~\ref{fig: gain_vs_R}). This suggests that the probability of gas breakdown might be mainly influenced by the quality (material and finish) of the electrodes' surfaces, and not much by their resistivity.

\begin{figure}
	\centering 	
 \includegraphics[width=0.5\textwidth]{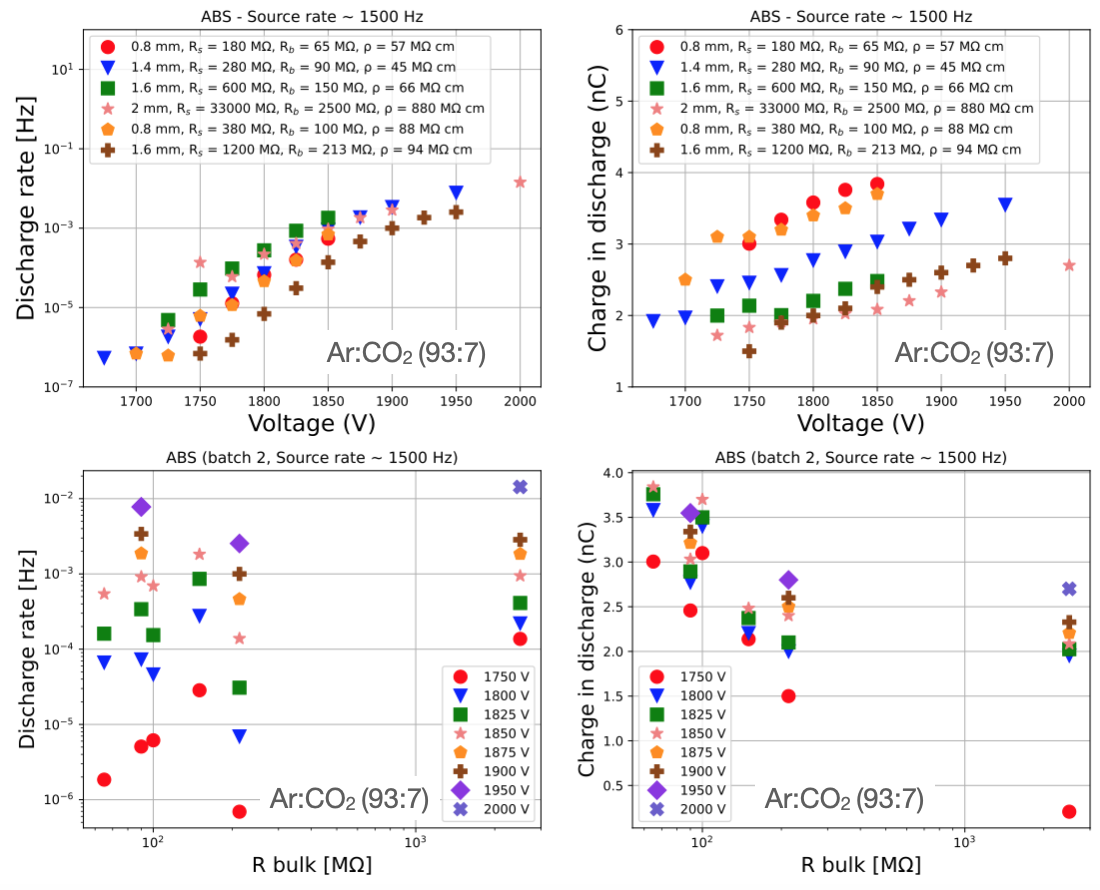}	
	\caption{Left: quenched discharge rate as a function of voltage (top) and of R$_b$ (bottom). Right: average quenched discharge intensity as a function of voltage (top) and of R$_b$ (bottom).}
	\label{fig: quenched-discharges}%
\end{figure}

Although smaller in magnitude, quenched discharges could still cause dead time or damage sensitive readout electronics. We characterized their intensity and rate by measuring induced signals on the anode. 
Figure~\ref{fig: quenched-discharges}-left shows the quenched-discharge rate as a function of voltage (top) and of R$_b$ (bottom). As for non-quenched discharges, no dependency on resistivity and thickness is observed. Figure~\ref{fig: quenched-discharges}-right shows the average quenched-discharge intensity as a function of voltage (top) and of R$_b$ (bottom). The intensity mildly increase with voltage, which might be due to pileup of multiple discharges or related to an increased charge mobility in the gas. A decrease in average discharge intensity was observed for increasing R$_b$ and R$_s$ as well as for increasing thickness. The latter might be due to a reduced signal induction related to a smaller weighting field. This effect should be further investigated.

\section{Charge equilibrium restoration in the resistive plate}

\begin{figure}
	\centering 	
 \includegraphics[width=0.5\textwidth]{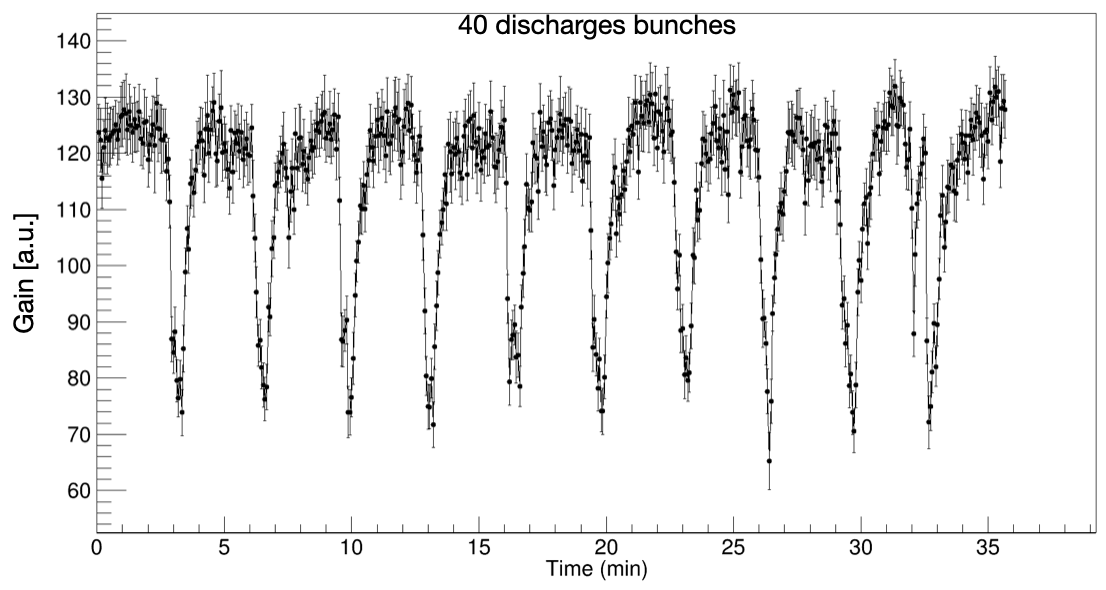}	
	\caption{Gain evolution 6~mm away from the discharge place, during bunches of 40 discharges.}
	\label{fig: gain_drop}%
\end{figure}

The effect of quenched discharges on the gain of an RPWELL detector was characterized with the same setup and methodology described in~\cite{jash2023electrical}. The RPWELL in this case included a semiconductive glass plate (see figure~\ref{fig: plates characterization}). Discharges were produced in a specific position for a fixed amount of time (at 0.5~Hz - 3~Hz rate), and the gain was monitored at different distances from the discharge position. Several consecutive discharges are needed to cause a measurable gain variation.  In figure~\ref{fig: gain_drop}, we show an example of gain evolution 6~mm away from the discharge place, during bunches of 40 discharges.
Here we focus on determining the typical recovery time of the detector. That is, the time it takes to neutralize the excess of electrons accumulated on the resistive plate after a discharge and restore the original gain value --  i.e, "charge evacuation time". For each discharge bunch, we fit the gain recovery curve to an exponential $y = B - e^{(x-x_0)/\tau}$, where $\tau$ is the charge evacuation time. 

\begin{figure}
	\centering 	
 \includegraphics[width=0.48\textwidth]{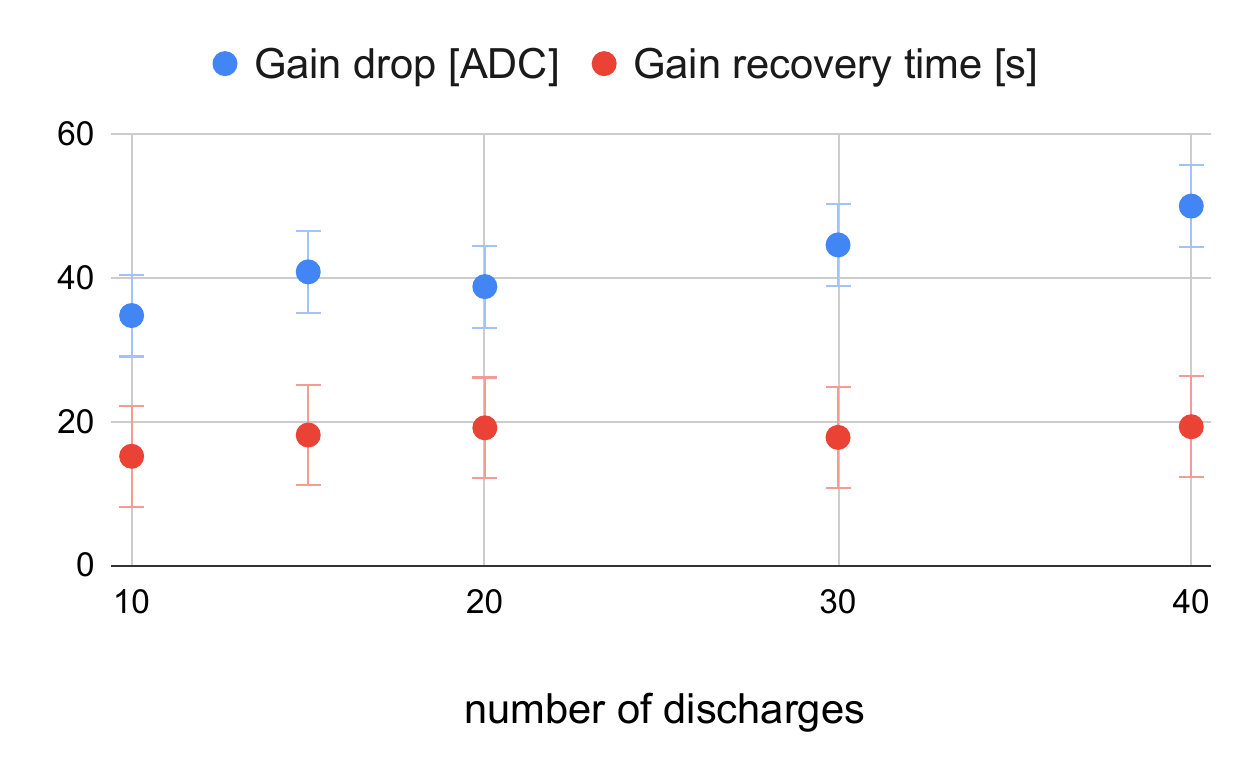}	
	\caption{Average gain drop (blue) and average gain recovery time (red) 6~mm away from the discharge place, as a function of the number of discharges in the bunch.}
	\label{fig: recovery_time}%
\end{figure}

Figure~\ref{fig: recovery_time} presents the average gain drop (left) and the average charge evacuation time (right) 6~mm away from the discharge place, as a function of the number of discharges in the bunch.
As can be seen, in the regime of 10 to 40 discharges per bunch, the gain drop is only mildly dependent on the number of discharges in the bunch. Moreover, the gain recovery time is independent on the number of discharges. 
In all cases the charge evacuation time is $\sim$20~s, which is about 3 orders of magnitude longer than the time estimated by approximating the RPWELL detector to an RC circuit (see~\cite{rubin2013first}). In that case the evacuation time would be $\tau = \rho \cdot \epsilon$, which for the semiconductive glass is in the ms scale. This hints to additional processes extending a naive capacitor charge-discharge model. The scale of seconds is more typical to diffusion of charges in static-dissipative materials~\cite{CHUBB1993273,sonnonstine1975surface}. This effect is important when evaluating and predicting the rate capability of detectors including resistive electrodes and it deserves further studies in the context of charge mobility in (semi-)insulating materials.

\section{Resistive layer of variable resistivity in Argon vapor}

In this section, we describe a qualitative characterization of discharge quenching in a cryogenic RWELL as a function of the anode surface resistivity in cryogenic conditions in Argon vapor. The resistive electrode consisted in a thin deposition of Diamond-Like Carbon (DLC~\cite{leardini2023diamond}) onto a Kapton foil. The surface resistivity of this material continuously increases with decreasing temperature. 
Using the same setup and configuration described in~\cite{LEARDINI2023168104}, we monitored the current from the power supply channels while slowly warming up the system from 90~$^{\circ}$K up to room temperature (293~$^{\circ}$K). To account for changes in the Argon vapor density during the warming up, the high voltage supplied to the RPWELL was constantly tuned to remain in the discharge regime.

\begin{figure}
	\centering 	
 \includegraphics[width=0.45\textwidth]{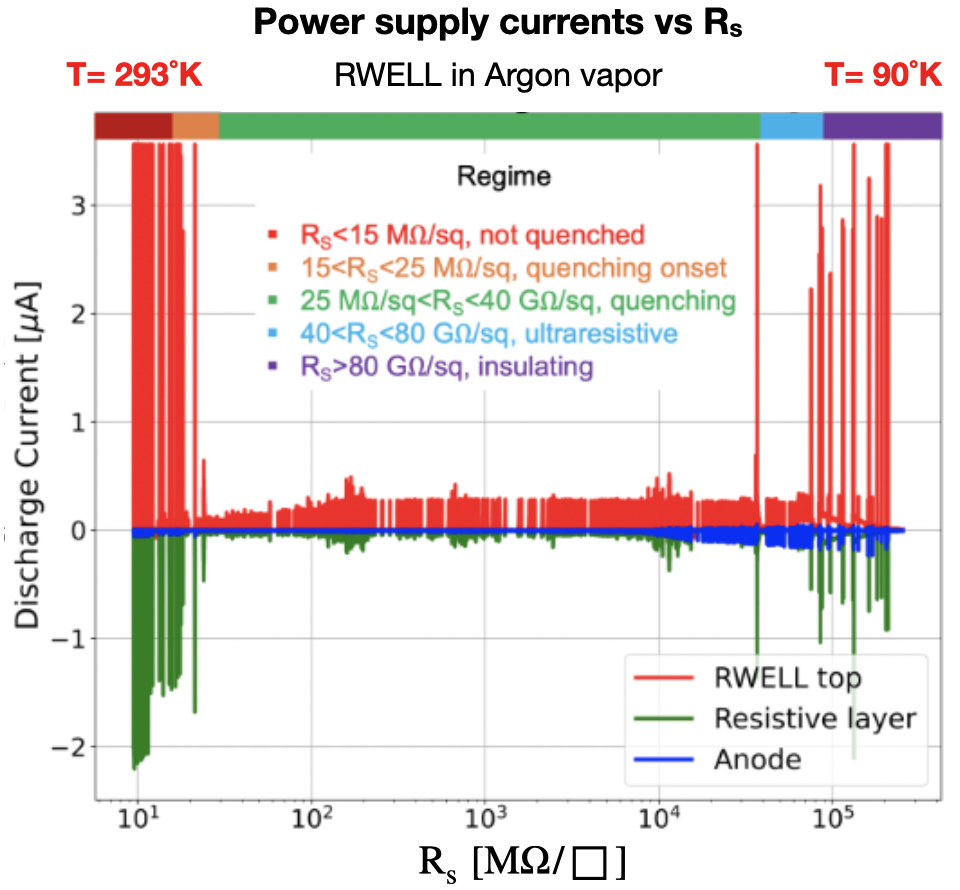}	
	\caption{Current from RWELL electrodes operated in cryogenic Argon vapor during warming up. The voltages were tuned to keep the detector in discharge regime. The resistivity of the resistive layer varies smoothly with the changing temperature.}
	\label{fig: cryogenic}%
\end{figure}

Figure~\ref{fig: cryogenic} shows the current values while the resistivity of the layer varied in the range of 10~M$\Omega/\square <$ R$_s <$ 10$^5$~M$\Omega/\square$~\cite{leardini2023diamond}.
Generally, in an RWELL detector the discharge quenching is less effective than in an RPWELL, therefore the discharges are measurable from the power supply even in the quenching regime. Five different regions are observed. For R$_s <$ 15~M$\Omega/\square$, there is no quenching effect. A sharp transition to quenching regime is observed at 15~M$\Omega/\square$. The discharge intensity varies until it becomes stable in the wide range 25~M$\Omega/\square <$
R$_s <$ 40~G$\Omega/\square$. Along this range, the discharges affect, as expected, only the RWELL top electrode and the resistive layer.
In the region 20~G$\Omega/\square <$
R$_s < 80$~~G$\Omega/\square$ the discharges are still quenched, but also the anode is affected by currents. We attribute this to a large charge evacuation time from the resistive layer inducing currents on the anode. 
For R$_s > 80$~~G$\Omega/\square$ the resistive layer behaves practically as an insulator resulting in large discharges breaking through the resistive layer.

This measurement can be considered only qualitatively, since different parameters are changing at the same time, some of them in a non-controlled way: vapor density, layer resistivity, detector gain, voltages. 
Nevertheless, for the first time we were able to show the discharge regime transitions when varying the resistivity of the resistive layer in a continuous way. To confirm these results, we repeated this experiment in the quenching regime while monitoring the discharge intensity by recording the scintillation light with a photomultiplier. Similar behaviour was observed. 

\section{Main findings and open questions}

Any tested RPWELL had a reduced discharge probability with respect to THWELL. Moreover, in the quenched regime, the discharge probability and intensity seem not to depend significantly on the resistance values of the resistive plate.
The charge evacuation time was measured to be about 3 orders of magnitude larger than what expected by a simplified RC model. 

\section*{Acknowledgements}
We would like to thank Dr. Yariv Pinto and Matan Divald from the Hebrew University of Jerusalem for the production of the ABS plates and prof. Yi Zhou (USTC) for the production of the DLC layers. We acknowledge the work of our colleagues at the Weizmann Institute of science: Dr. Gregory Leitus for his assistance with electrical characterization of the plates and Dr. Xiaomeng Sui for the SEM images. 
\\Thanks to Ryan Felkai for the cryogenic measurements of scintillating light.
\\This work was supported by Grant No. 3177/19 from the Israeli Science Foundation (ISF), The Pazy Foundation, and by the Sir Charles Clore Prize.

%\bibliographystyle{elsarticle-harv} 
\bibliographystyle{elsarticle-num} \bibliography{bibliography.bib}

\end{document}